\documentclass[twocolumn, superscriptaddress, showpacs,
 amsmath,amssymb,
 aps,
 prab
]{revtex4-2}
\usepackage{graphicx} 
\usepackage{xcolor}

\newcommand{\m}{\text{m}}
\newcommand{\kev}{\text{keV}}
\newcommand{\mev}{\text{MeV}}
\newcommand{\gev}{\text{GeV}}
\newcommand{\tev}{\text{TeV}}

\newcommand{\cm}{\text{cm}}

\newcommand{\mum}{\text{$\mu$}\m}

\newcommand{\joule}{\text{J}}
\newcommand{\s}{\text{s}}
\newcommand{\fs}{\text{fs}}
\newcommand{\ns}{\text{ns}}
\newcommand{\ms}{\text{ms}}

\newcommand{\pc}{\text{pC}}

\newcommand{\mrad}{\text{mrad}}
\newcommand{\sr}{\text{sr}}

\newcommand{\g}{\text{g}}
\newcommand{\mus}{\text{$\mu$\s}}
\newcommand{\hz}{\text{Hz}}

\begin{document}

\preprint{APS/123-QED}

\title{Measurement of directional muon beams generated at the Berkeley Lab Laser Accelerator}

\author{Davide Terzani}
\email{dterzani@lbl.gov}
\affiliation{%
 Lawrence Berkeley National Laboratory, Berkeley, CA 94720, USA
}
\author{Stanimir Kisyov}
\affiliation{%
 Lawrence Berkeley National Laboratory, Berkeley, CA 94720, USA
}
\author{Stephen Greenberg}
\affiliation{%
 Lawrence Berkeley National Laboratory, Berkeley, CA 94720, USA
}
\affiliation{Department of Physics, University of California, Berkeley, CA 94720, USA}
\author{Luc Le Pottier}
\affiliation{%
 Lawrence Berkeley National Laboratory, Berkeley, CA 94720, USA
}
\affiliation{Department of Physics, University of California, Berkeley, CA 94720, USA}
\author{Maria Mironova}
\affiliation{%
 Lawrence Berkeley National Laboratory, Berkeley, CA 94720, USA
}
\author{Alex Picksley}
\affiliation{%
 Lawrence Berkeley National Laboratory, Berkeley, CA 94720, USA
}
\author{Joshua Stackhouse}
\affiliation{%
 Lawrence Berkeley National Laboratory, Berkeley, CA 94720, USA
}
\affiliation{Department of Nuclear Engineering, University of California, Berkeley, CA 94720, USA}
\author{Hai-En Tsai}
\affiliation{%
 Lawrence Berkeley National Laboratory, Berkeley, CA 94720, USA
}
\author{Raymond Li}
\affiliation{%
 Lawrence Berkeley National Laboratory, Berkeley, CA 94720, USA
}
\affiliation{Department of Nuclear Engineering, University of California, Berkeley, CA 94720, USA}
\author{Ela Rockafellow}
\affiliation{%
 Institute for Research in Electronics and Applied Physics and Department of Physics, University of Maryland, College Park, Maryland 20742, USA
}
\author{Bo Miao}
\affiliation{%
 Institute for Research in Electronics and Applied Physics and Department of Physics, University of Maryland, College Park, Maryland 20742, USA
}
\author{Jaron E. Shrock}
\affiliation{%
 Institute for Research in Electronics and Applied Physics and Department of Physics, University of Maryland, College Park, Maryland 20742, USA
}
\author{Timon Heim}
\affiliation{%
 Lawrence Berkeley National Laboratory, Berkeley, CA 94720, USA
}
\author{Maurice Garcia-Sciveres}
\affiliation{%
 Lawrence Berkeley National Laboratory, Berkeley, CA 94720, USA
}
\author{Carlo Benedetti}
\affiliation{%
 Lawrence Berkeley National Laboratory, Berkeley, CA 94720, USA
}
\author{John Valentine}
\affiliation{%
 Lawrence Berkeley National Laboratory, Berkeley, CA 94720, USA
}
\author{Howard M. Milchberg}
\affiliation{%
 Institute for Research in Electronics and Applied Physics and Department of Physics, University of Maryland, College Park, Maryland 20742, USA
}
\author{Kei Nakamura}
\affiliation{%
 Lawrence Berkeley National Laboratory, Berkeley, CA 94720, USA
}
\author{Anthony J. Gonsalves}
\affiliation{%
 Lawrence Berkeley National Laboratory, Berkeley, CA 94720, USA
}
\author{Jeroen van Tilborg}
\affiliation{%
 Lawrence Berkeley National Laboratory, Berkeley, CA 94720, USA
}
\author{Carl B. Schroeder}
\affiliation{%
 Lawrence Berkeley National Laboratory, Berkeley, CA 94720, USA
}
\affiliation{Department of Nuclear Engineering, University of California, Berkeley, CA 94720, USA}
\author{Eric Esarey}
\affiliation{%
 Lawrence Berkeley National Laboratory, Berkeley, CA 94720, USA
}
\author{Cameron G. R. Geddes}
\affiliation{%
 Lawrence Berkeley National Laboratory, Berkeley, CA 94720, USA
}
\date{\today}

\begin{abstract}
We present the detection of directional muon beams produced using a PW laser facility at the Lawrence Berkeley National Laboratory.
The muon source is a multi-GeV electron beam generated in a $30\,\cm$ laser plasma accelerator interacting with a high-$Z$ converter target.
The GeV photons resulting from the interaction are converted into a high-flux, directional muon beam via pair production.
By employing scintillators to capture delayed events, we were able to identify the produced muons and characterize the source.
Using theoretical knowledge of the muon production process combined with simulations that are in excellent agreement with the experiments,
we demonstrate that laser-plasma accelerators have the capability of generating electron beams with characteristics suitable to produce $\gev$-scale muons
that offer unique advantages with respect to the cosmic background.
Laser-plasma-accelerator-based muon sources can therefore enhance muon imaging applications thanks to their compactness, directionality, and high yields,
which reduce the exposure time by orders of magnitude compared to cosmic ray muons.
Using the Geant4-based simulation code we developed to gain insight into the experimental results,
we can design future experiments and applications based on LPA-generated muons.
\end{abstract}

\maketitle
\section{Introduction}

Muons beams have attracted considerable attention in recent years because of their unique properties and potential applications across various fields of physics.
However, production of muon beams for fundamental physics research,
for instance a future TeV-class muon-muon collider~\cite{schulte_muon_2022} or the Muon g-2 experiment at Fermilab~\cite{keshavarzi_muon_2022},
is challenging and it is currently done by accelerating high-energy
proton beams in large-scale facilities, colliding them with fixed targets and
collecting the created mesons that will eventually decay into muons~\cite{zisman_proton_2010}.

Given the large size of the accelerators,
proton-based muon production is not suitable for imaging applications
because neither the source nor the typical imaging samples can be transported.
In fact, thanks to their considerable penetration power,
muons are used for nondestructive imaging of large or concealed objects~\cite{bouteille_micromegas-based_2016}, which must be performed on-site.
For such reason, currently the only available source of muons for imaging applications are the cosmic rays,
which, when interacting with the Earth's atmosphere, generate showers of particles, including muons, that are accessible in any unshielded place on the planet.
The incoming muon flux on the Earth's surface can be approximated as $dN/d\Omega dS dt\approx70\cos^2\left(\theta\right)\m^{-2}\s^{-1}\sr^{-1}$~\cite{particle_data_group_review_2022},
where $\theta$ is the angle from the zenith, along which the flux is maximum.
For most of the imaging applications, such a muon flux requires
months of exposure before a statistically significant signal is accumulated.
For instance, when imaging objects along the plane of earth's surface, the muon flux is reduced
to $dN/dS dt \ll 0.1 - 1\, \m^{-2}\s^{-1}$ due to the limited angular aperture of the detectors.
Therefore, muography via cosmic ray muons can generally be applied only to objects that are
immobile over long periods of time.
A notable and fascinating example of their use was the discovery of a hidden chamber
in Khufu's Pyramid (Great Pyramid of Giza)~\cite{morishima_discovery_2017, nishio_development_2015}.
They also enable investigating the magmatic chambers of active volcanoes~\cite{tanaka_development_2003},
blast furnaces~\cite{hu_exploring_2018}, and nuclear waste~\cite{fujii_performance_2013, fujii_detection_2017, jonkmans_nuclear_2013}.
These are only a few examples of the available applications and
implementations of the muography technique.
For a more general review see Refs.~\cite{vanini_muography_2018, procureur_muon_2018} and references therein.

Laser plasma accelerators (LPAs) provide an alternative source of muon beams as a
byproducts of the collision of an LPA-generated electron beam with a high-$Z$ target~\cite{titov_dimuon_2009},
a possibility supported by some numerical studies~\cite{rao_bright_2018, calvin_laser-driven_2023}.
Several laboratories have demonstrated generation of $100s\,\mev$ to multi-GeV-class electron beams in targets up to tens of centimeters using different techniques, with representative examples in Refs.~\cite{aniculaesei_acceleration_2023, faure_laserplasma_2004, geddes_high-quality_2004, leemans_multi-gev_2014, mangles_monoenergetic_2004, wang_quasi-monoenergetic_2013, gonsalves_petawatt_2019, picksley_matched_2024, shrock_guided_2024,*rockafellow_development_2025, *miao_multi-gev_2022},
providing enough energy for muon generation via pair-production.
Pair-produced muons are very suitable for muography as they are characterized by high-energy,
of the order of the initial electron energy, and low divergence and they are generated directly into
the solid target where the interaction is taking place.
Currently available high-power laser systems, operating at repetition rates between 1 and 10 Hz,
enable producing LPA-based directional muon beams with yields orders of magnitude higher than cosmic rays,
thus providing substantial speedups to applications that typically require months of exposure times.
A next generation kHz LPA~\cite{kiani_high_2023} could further improve this factor by three orders of magnitude,
opening new avenues to the use of muons in imaging large objects in only a few minutes.
A LPA-based muon source provides the essential characteristics that are needed for imaging applications together with the advantages of a compact source that can
be deployed on the imaging site.
Laser technologies currently in development, such as fiber lasers or high-efficiency solid state lasers, show the potential of realizing such a compact and deployable laser system in the future~\cite{kiani_high_2023}.

In this paper we present the generation and measurement of GeV muons using the BErkeley Laboratory Laser Accelerator (BELLA) Facility at the Lawrence Berkeley National Laboratory.
In Sec.~\ref{sec:theory} we give an overview of the relevant muon production mechanisms,
highlighting the characteristics of the generated muons.
In Sec.~\ref{sec:experimental_setup} we present the setup for the electron beam generation
and the production and measurements of muons.
In Sec.~\ref{sec:experimental_results} we analyze the results of the experiments and show
how muons are identified.
We demonstrate that the electron beams interacting with the shielding configuration of our laboratory
generate muons via different mechanism and that we can spatially separate them into a directional
and a quasi-isotropic source.
Simulations that reproduce the experimental measurements are presented in Sec.~\ref{sec:simulation_results},
where we investigate in detail the muon production process in the experimental cave.
Sec.~\ref{sec:conclusions} presents the conclusions of our work and the next steps.

\section{Mechanisms of muon production in high-$Z$ materials}\label{sec:theory}
When a high-energy electron traveling in a solid converter target is deviated by an atomic nucleus,
it emits high energy photons, or \emph{Bremsstrahlung} radiation~\cite{jackson_classical_1962}.
The energy distribution of the Bremsstrahlung photons is characterized by an exponential-like decay,
extending to the same maximum energy as the initial electron~\cite{jackson_classical_1962, chao_handbook_2013, tsai_pair_1974}.
The angle of emission is confined within a cone of angular aperture $\Delta \theta_B \simeq 1/\gamma_e$,
where $\gamma_e$ is the Lorentz factor of the radiating electron.
Bremsstrahlung photons that interact with the background atomic nuclei can be converted into pairs of charged leptons
(a particle and an antiparticle) via the \emph{Bethe-Heitler} process if their energy is $E_\gamma\geq 2m_pc^2$,
where $m_p$ is the particle (lepton) mass, and $c$ the speed of light in vacuum.
The cross-section for the Bethe-Heitler production is proportional to the inverse of the particle mass squared~\cite{bethe_stopping_1997},
i.e., $\sigma_{BH}\propto 1/m_p^2$, and the generated particles are emitted along the photon propagation direction,
thus resulting in a particle shower confined within an angle $\lesssim \Delta \theta_B$.
The energy spectrum of the products follows the Bremsstrahlung distribution,
with a maximum energy $E_{max}=E_\gamma-2m_p c^2$, due to energy-momentum conservation.

Given that $e^+e^-$ pairs, i.e., positron-electron, are the most probable products of the Bethe-Heitler process,
Bremsstrahlung radiation induced by the propagation of an electron beam into high-$Z$ targets
(e.g., tungsten or lead) has been used as a source of positron beams
for conventional acceleration for several decades~\cite{chao_handbook_2013}
and it has been extensively investigated in experiments.

Both the Bremsstrahlung and the Bethe-Heitler emissions develop over a characteristic radiation length $X_0$.
The radiation length is defined as the length over which an electron propagating in the material loses all but $1/e$ of its initial energy, which corresponds to $7/9$ of the mean free path for Bethe-Heitler emission.
The value of $X_0$ depends on the atomic composition of the material and on its density.
For high-$Z$ materials, its value ranges from millimeters to centimeters~\cite{chao_handbook_2013}.
\begin{figure}[ht]
\centering
\includegraphics[width=8.6cm]{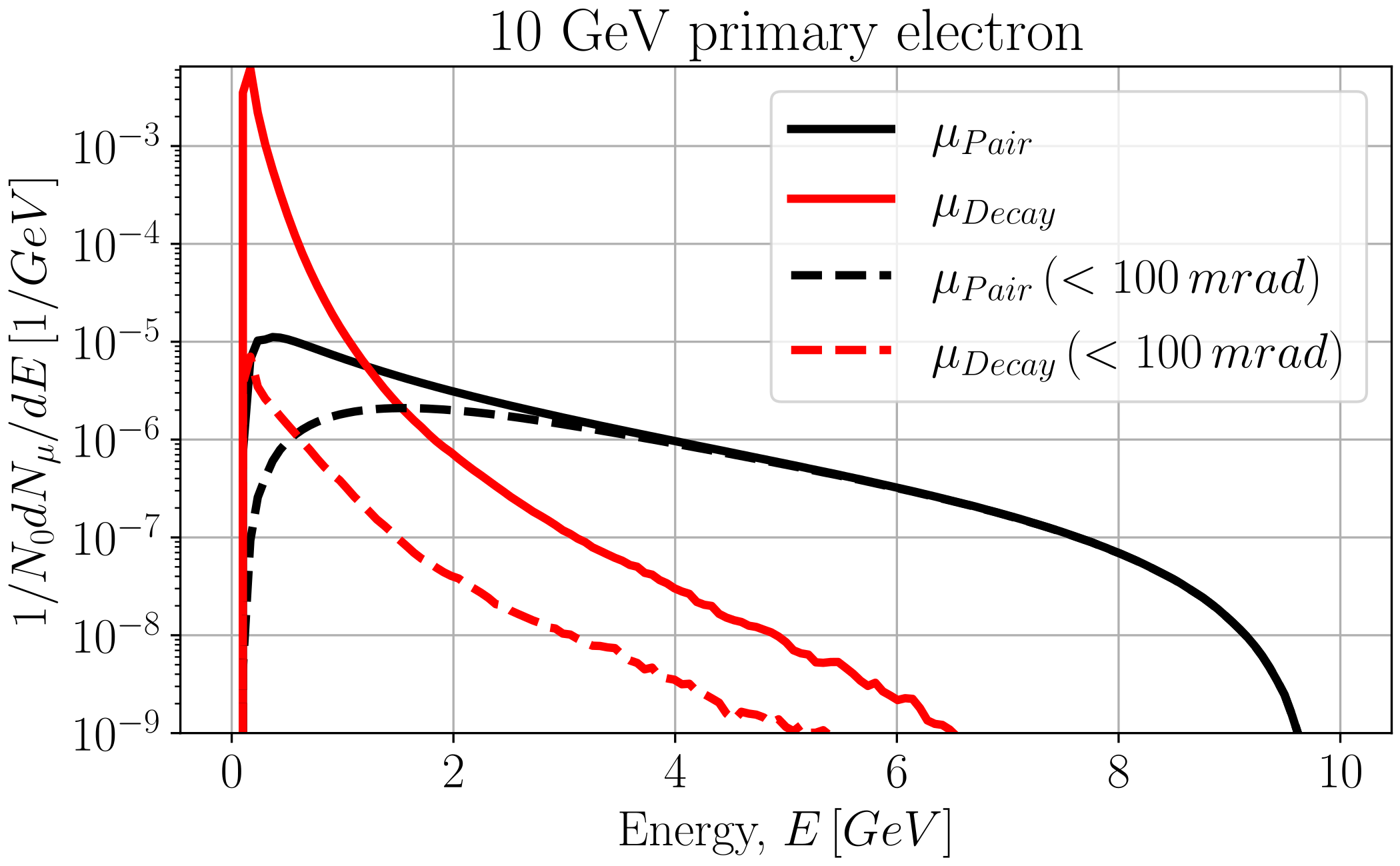}
\caption{
Simulated energy spectra of muons generated via pair production and decay from a $10\,\gev$ electron beam propagating in a $2.1\,\cm$-thick tungsten target.
The distributions are normalized per initial primary electron.
The solid black and red lines correspond to the total yield of pair produced and decay muons, respectively.
The dashed black and red lines present the spectrum of respectively pair and decay produced muons within $100\,\mrad$.
Simulation results show that effectively all pair produced muons with $E\simeq 4\,\gev$ are confined within $100\,\mrad$.}
\label{fig:muon_spectrum}
\end{figure}

The properties of photon decay into $\mu^+\mu^-$ pairs are analogous to the $e^+e^-$ pair creation,
but the cross section is reduced by a factor 
$R=m_e^2/m_\mu^2=2.34\times 10^{-5}$.
For this reason, it has not been subject to the same detailed experimental investigations and applications.

In addition to the Bethe-Heitler mechanism,
the interactions of electrons with targets also lead to the production of pions ({$\pi^\pm$}) which subsequently decay into muons
(the average pion lifetime is $\tau_{\pi^\pm}=2.6\times10^{-8}\,\s$).
In particular, main sources of pions in the material are the processes
$\gamma + p \rightarrow \pi^{+} + n$,
$\gamma + n \rightarrow \pi^{-} + p$,
and $\gamma + N \rightarrow \pi^{+} + \pi^{-} + N$,
where the $\gamma$ rays interacting with the background protons ($p$), neutrons ($n$), and nucleons ($N$\,=\,$p$, $n$)
are the ones produced via Bremsstrahlung~\cite{titov_dimuon_2009}.
Analogous processes apply to kaon creation, although with a lower probability due to the higher kaon mass.
Both pions and kaons decay into muons,
therefore we will refer to the muon flux resulting from this mechanism as coming from a more general meson decay.
The total cross section for muon production through meson decay in a high-$Z$ target is higher than the one for the Bethe-Heitler $\mu^{+}\mu^{-}$ process.
However, given the large meson rescattering and absorption cross sections,
the resulting meson flux is weaker and not well collimated in a thick target~\cite{titov_dimuon_2009}.
Therefore, we expect some contributions to the number of muons from meson decay at large angles.
Moreover, due to the relativistic dilation of the proper time, only low energy mesons decay in the vicinity of the converter,
resulting in a sharp cut-off of the spectrum at high energies.
It is possible to produce high energy muons via meson decay, but it requires $\sim$km-scale separations from the converter.

Fig.~\ref{fig:muon_spectrum} shows simulated energy spectra, normalized per initial primary electron,
of muons generated respectively via pair production and meson decay from a monochromatic $10\,\gev$ electron beam
propagating in a $2.1\,\cm$-thick tungsten target.
The simulations were performed using the Geant4 toolkit for simulation of the propagation of particles through matter~\cite{agostinelli_geant4simulation_2003,allison_geant4_2006, allison_recent_2016},
including solely the monoenergetic electron beam and tungsten target in the setup geometry.
The solid black line is the spectrum of muons generated via Bethe-Heitler integrated over $4\pi$,
while the dashed black line is the spectrum of muons selected within a $100\,\mrad$ cone,
which effectively contains all the muons produced with energies $E\gtrsim4\,\gev$.
The red line is the energy spectrum of muons generated via meson decay over the $4\pi$ angle.
Muons from meson decay with energies $\geq 1\,\gev$ are statistically negligible within $100\,\mrad$, as shown by the dashed red line.
\begin{figure}[ht]
\centering
\includegraphics[width=8.6cm]{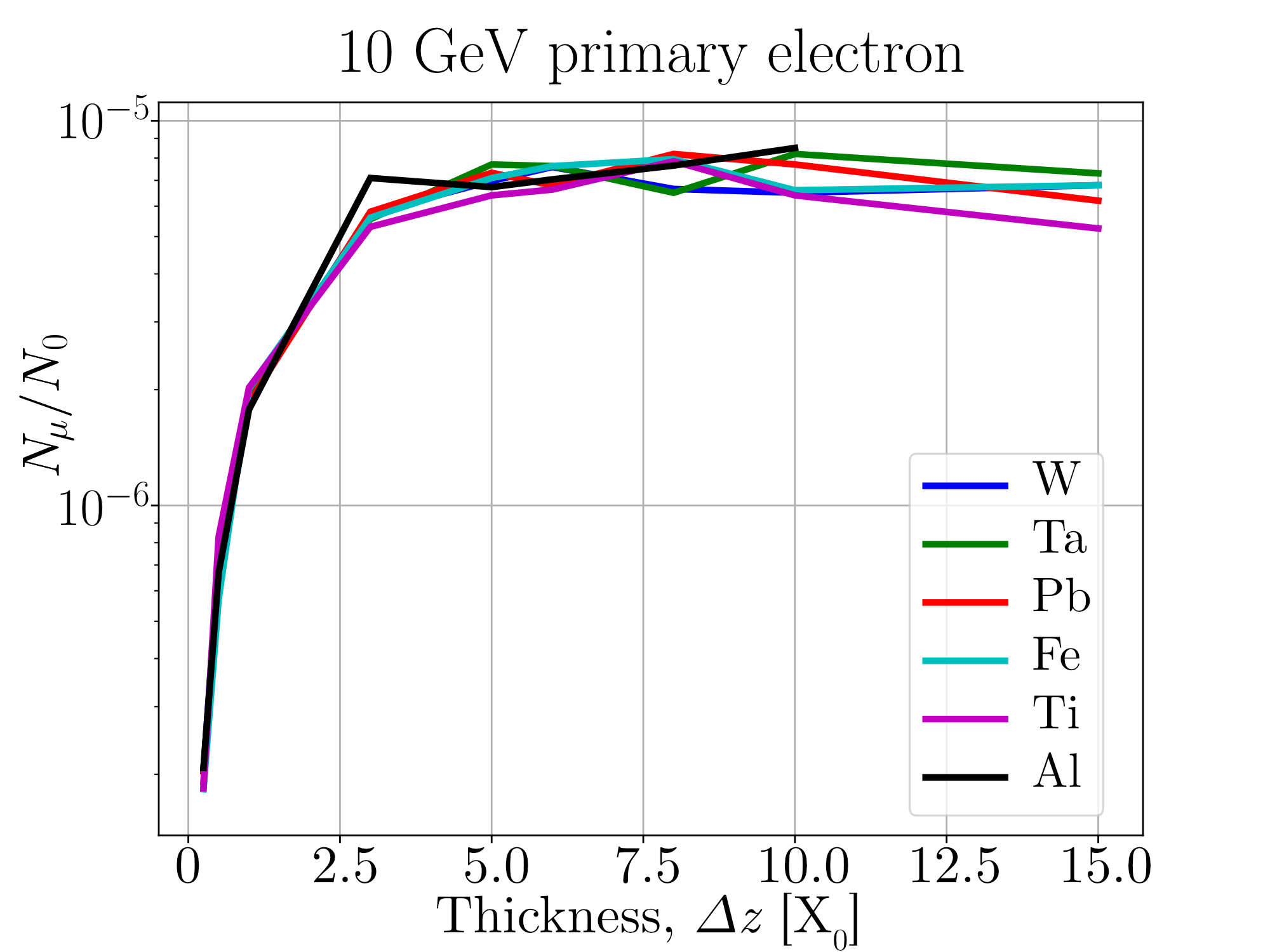}
\caption{
Simulated yields of pair produced muons within $100\,\mrad$ along the primary $10\,\gev$ electron beam axis for different materials and their thicknesses expressed in radiation lengths $X_{0}$.
Normalization to a single primary electron is applied.
The plot shows a saturation of the yield at $\sim 6\,X_{0}$ and independence of the target material.
Beyond $10\,X_0$, muon production is substantially suppressed,
and ionization energy losses prevent low-energy muons from escaping the target. This leads to a reduction in the observed yield, reflecting the decreasing contribution of low-energy muons to the total yield.}
\label{fig:material_scan}
\end{figure}

We show the effect of target thickness on the number of muons produced per initial electron in Fig.~\ref{fig:material_scan}, with results from Geant4 simulations.
The pair production only depends on the normalized thickness,
i.e., the number of radiation lengths traversed,
and does not show significant differences between different materials.
The number of pair-produced muons is maximum between $(6-10)\,X_0$ for an initial $10\,\gev$ electron beam,
similarly to what is expected for positrons~\cite{chao_handbook_2013},
and slowly decreases for larger thicknesses due to the high penetration power of muons.
We point out that the processes discussed here are single-particle interactions and, therefore, the number of muons produced is linearly proportional to the incoming electron beam charge.

In this section we reviewed different mechanisms for muon production,
highlighting their characteristics.
Pair production yields a directional and collimated muon beam,
while meson decay generates a quasi-isotropic muon flux.
In the next sections we will show how the interaction of an electron beam with the shielding
configuration of our laboratory combines these processes.
The electron beam dump suppresses the meson propagation, absorbing all but muon produced via pair production,
while mesons are generated in other high-$Z$ elements along the beam path,
which do not possess the same filtering power,
and eventually decay into a quasi-isotropic background of low-energy muons.

\section{Experimental setup}\label{sec:experimental_setup}
\begin{figure*}[ht]
\centering
\includegraphics[width=17.2cm]{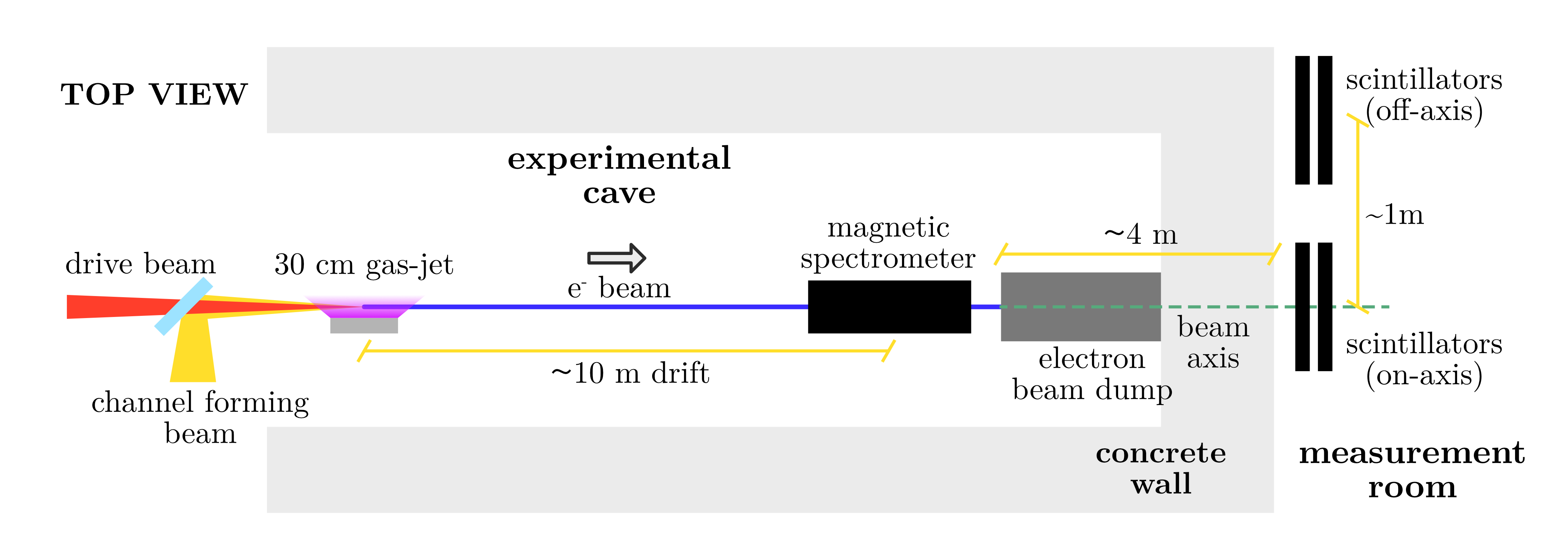}
\caption{
Schematic top-view of the experimental setup (not to scale).
The laser drive and channel-forming beams, on the left, are coupled into a $30\,\cm$ gas-jet, generating the electron beam.
After a $\sim10\,\m$ drift, the electron beam reaches the magnetic spectrometer and is then unloaded on the electron beam dump,
where it produces the muons.
The scintillators are placed in the measurement room, behind the concrete wall, in two possible configurations.
Each configuration includes a pair of scintillators close to each other (spaced a few cm),
which were placed either along the beam axis, i.e., ``on-axis'' configuration,
or $\simeq 1\,\m$ to the side of the beam axis, ``off-axis'' configuration.
The scintillators are $\simeq 10\,\cm$ away behind the concrete wall.
}
\label{fig:experiment_scheme}
\end{figure*}
The experiments described here were carried out in an experimental campaign to generate and measure muons at the BELLA center between December 2023 and January 2024.
The laser-plasma accelerator was driven using the BELLA PW laser~\cite{nakamura_diagnostics_2017}.
A detailed description of the accelerator can be found in~\cite{picksley_matched_2024}
and a schematic of the experiment can be found in Fig.~\ref{fig:experiment_scheme}.
The drive laser duration was $\sim 40\,\fs$ and was focused to a spot-size of $w_0 = (53 \pm 1 )\, \mum$, at the entrance of a 30-cm-long hydrogen gas jet~\cite{picksley_matched_2024, miao_meter-scale_2025}.
The measured laser energy was $E\simeq 21\,\joule$.
Before the arrival of the LPA drive laser, an auxiliary pulse focused by an axicon lens was used
to ionize and form a plasma channel by hydrodynamic expansion~\cite{durfee_light_1993} of an optical field-ionized plasma~\cite{shalloo_hydrodynamic_2018, *shalloo_low-density_2019, *picksley_meter-scale_2020, *picksley_guiding_2020, *picksley_all-optical_2023, lemos_guiding_2018, feder_self-waveguiding_2020, *miao_optical_2020}.
The measured axial density was $n_0 \approx 1 \times 10^{17}\cm^{-3}$.
Ionization injection was triggered by a 1\% nitrogen dopant extending over a region of the channel of length $L_\mathrm{dop}$. For $L_\mathrm{dop} \approx 12\,\cm$, it was possible to achieve single, quasimonoenergetic electron bunches with peak energy up
to $9.2\,\gev$ and charge extending beyond $10\,\gev$~\cite{picksley_matched_2024}.

\begin{figure}[ht]
\centering
\includegraphics[width=8.6cm]{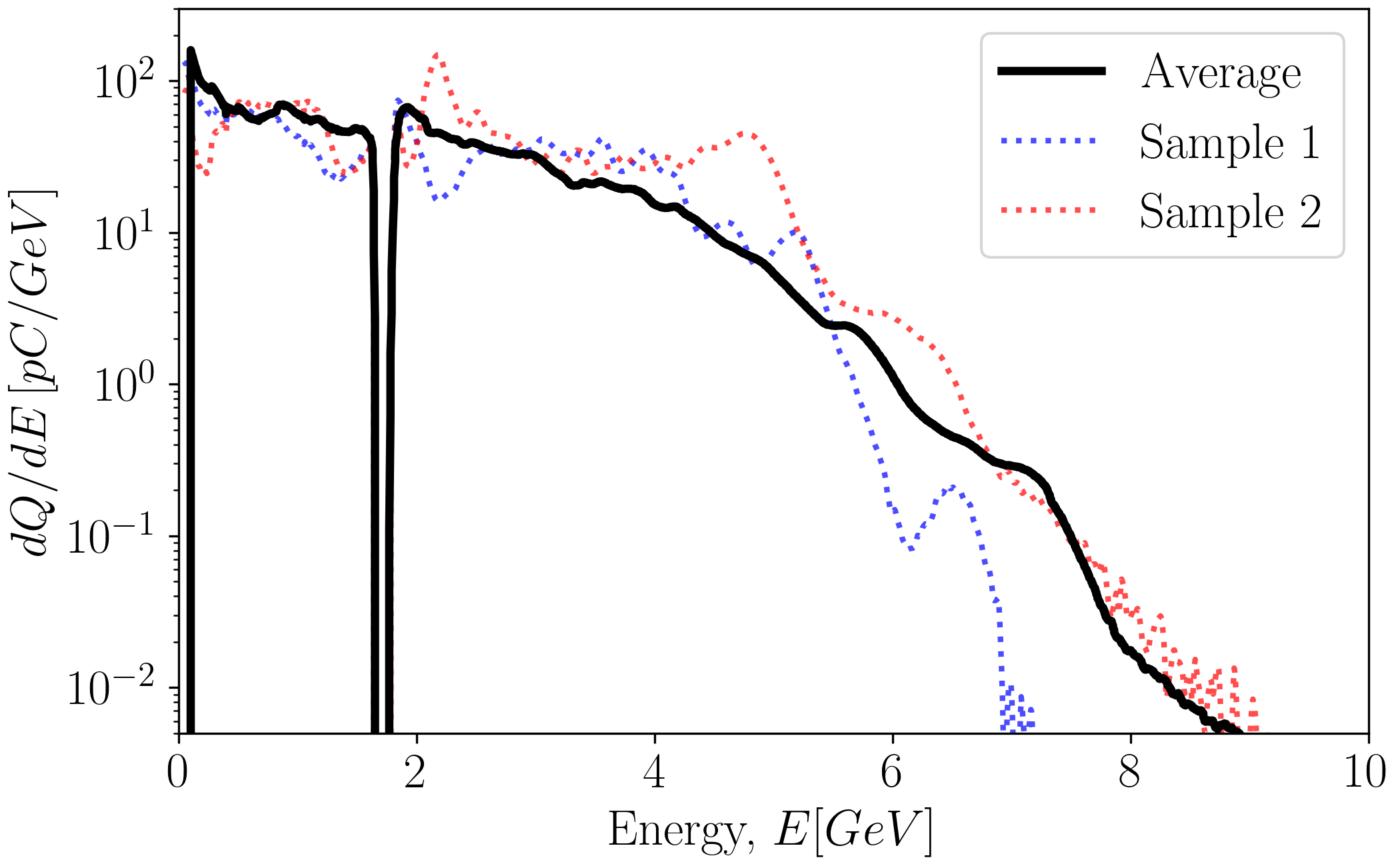}
\caption{The solid black line is the electron energy spectrum (averaged over 42 shots) used for muon generation experiments in December 2023.
The beam is characterized by continuous, exponential-like spectrum with a tail extending to $\sim 8\,\gev$.
Its integrated charge for energies $E\geq 2\,\gev$ is $Q\simeq80\,\pc$.
The blue and red dotted lines are two samples of single-shot electron spectra of the generated beams.
The dip around $1.9\,\gev$ is due to a physical gap in the magnetic spectrometer screens.}
\label{fig:electron_spectrum}
\end{figure}

For this experiment, $L_\mathrm{dop} \approx 30\,\cm$, and the accelerator was run in a stable condition with electron beam spectra show in Fig.~\ref{fig:electron_spectrum}. The measured charge was in the range $Q\simeq 45-250\,\pc$, with an average $Q\simeq 80\,\pc$, integrated over energies above $2\,\gev$,
and exponential-like energy spectra with tails extending to $\sim 8\,\gev$ were observed.

These high-energy electrons are passed through a $1.08\,\text{T}$ momentum analyzing dipole magnet (which deflects the electron beam downward towards the floor). We specifically refer to the 4GeV electron trajectory axis through and past the magnet as the ``deflected reference axis''. The deflected electrons are then stopped in a dedicated electron beam dump composed of $40.5\,\cm$ of lead, $1\,\m$ of steel, and a final $1.80\,\m$ block of concrete.
After the electron beam dump, a $90\,\cm$ concrete wall separates the experimental cave from the measurement room.
The measurement room, located directly behind the wall in the beam direction  propagation, is where the detectors were placed.
The length $L$ of the electron beam dump, expressed in units of radiation lengths, is $L\simeq 142$,
calculated using $X_{0,\text{Pb}}=0.56\,\cm$, $X_{0,\text{Steel}}=1.76\,\cm$, and $X_{0,\text{Concrete}}=11.6\,\cm$.
The large thickness suppresses the propagation of secondary particles, such as photons, pions, kaons, electrons, and positrons.
Muons only lose energy via ionization with a rate per unit density of about $2\,\mev \cm^2/\g$~\cite{groom_muon_2001}.
Energy losses via Bremsstrahlung are negligible and are only relevant for muons with energies $E\gtrsim 1\,\tev$.
This allows high energy muons to propagate through the beam dump and escape from its rear, pass through the wall, and finally reach the measurement room, where they can be detected.
From our simulations and from analytical estimates using the stopping power provided in Ref.~\cite{groom_muon_2001},
and considering the full beam dump and the wall,
we obtain that a muon loses about $E_L\lesssim 4\,\gev$ via ionization when reaching the measurement room.
Ionization losses act as a filter for the low energy muons, as muons generated with energies $E<E_L$ are absorbed during their propagation.
Energy losses via ionization are energy-independent in the energy range of interest for our experiment,
so the estimate holds for all the muons produced in the interaction.
These losses must be taken into account when designing a muon source as they typically define
the muon penetration range into the imaged sample.
Muons propagating in the material are subject to scattering, which increases the final muon beam divergence.
An estimate of the final angle accumulated by a muon of momentum $p$,
that is with normalized velocity $\beta=v/c=p/(m_\mu c\gamma)$,
via multiple scattering after propagating a distance $d$ in a material with radiation length $X_0$
can be obtained using the approximate formula~\cite{fruhwirth_quantitative_2001}
\begin{equation}
    \sigma^2_\theta=\left( \frac{13.6\,\mev}{p\beta}\right)^2\frac{d}{X_0}\left[1+0.038 \log\left(\frac{d}{\beta^2X_0}\right)\right]^2,
    \label{eq:moliere}
\end{equation}
derived from Moliere's scattering theory~\cite{bethe_molieres_1953}.
Eq.~\eqref{eq:moliere} neglects energy losses during the propagation,
but we can apply it using the average energy of a muon to roughly estimate the final angular aperture of the beam.
For instance, for muons with an initial energy $E_0=4\,\gev$ the final \emph{rms} divergence is on the order of $\sigma_\theta\simeq 100\,\mrad$.
Given a total propagation length from the creation point to the measurement room,
i.e., the beam dump and the wall, of about $L\simeq 4\,\m$,
the expected beam \emph{rms} size behind the wall is $\sigma_r\simeq 50\,\cm$.

\subsection{Detectors}

A scintillator-based radiation detector is employed for beam pulse monitoring
and muon identification behind the beam dump along the aforementioned deflected reference axis.
The setup consists of two panels of scintillating plastic ($1\,\cm$ thick Polystyrene,
total area of  $500\,\cm^2$/panel) each optically coupled to a photo-multiplier tube (PMT).
Each PMT signal is fed into a shaping amplifier and, along with a logic trigger indicating the laser pulse timing,
read out by a 50 MHz Analog-to-Digital Converter (ADC). A Nuclear Instrumentation Module logic unit giving the `AND' of both channel discriminators initiates a  $50\,\mus$
readout window of the ADC, which is saved for offline analysis.
The readout is activated when both scintillators detect a signal at approximately the same time,
for instance when they are crossed by the initial burst of particles generated in the electron beam interaction with the shielding,
which defines the $t=0$ time.
The trigger timestamp was subsequently verified during the post-processing analysis, matching it with the recorded timestamp of each LPA shot.
The time resolution of particle interaction in the scintillator is limited by the period between samples of the ADC, which is $20\,\ns$.

Fig.~\ref{fig:experiment_scheme} shows the experimental apparatus behind the beam dump.
The scintillator pair is arranged in two different orientations used to monitor the muon
production at different angles from the beam direction.
In particular, we identify an ``on-axis'' configuration,
in which the pair is along the deflected reference axis with the two scintillators separated along this axis by just a few cm,
and an ``off-axis'' configuration in which the pair of detectors is moved about $1 \,\m$ horizontally (sideways).
Note that at the scintillator location, the ``non-deflected reference axis'' for the particles
(in absence of a downward deflecting magnetic field) would be approximately $1\,\m$ higher,
but this scenario was not experimentally studied in this campaign.

Muons in the 1-10 GeV energy range are minimum ionizing particles, however, below this range,
their energy loss per unit distance rises sharply~\cite{groom_muon_2001}.
Muons (antimuons) of sufficiently low energy, on the order of $20\,\mev$,
can be stopped by the layers of plastic scintillator,
and their decay to an electron (positron) and two neutrinos within the plastic can subsequently be observed with an average lifetime $\tau_\mu=2.2\,\mus$.
Stopped muon decays are identified in the experimental apparatus via the ionization from the daughter electron,
which manifests as a second light pulse following the initial trigger.
By studying the decay time distribution, a positive muon identification can be made
by checking consistency with the muon lifetime.

\section{Experimental results}\label{sec:experimental_results}

During the experiment, we operated the laser at a frequency of $0.1\,\hz$.
The radiation generated by the accelerated electron beam unloaded on the electron beam dump
served as a trigger to activate the scintillators.
Of the muons produced in the interaction, some reach the detectors with an energy range $0\leq E\lesssim 20\,\mev$.
These muons stop in the Polystyrene until they decay, generating a delayed signal.
\begin{figure}[ht]
\centering
\includegraphics[width=8.6cm]{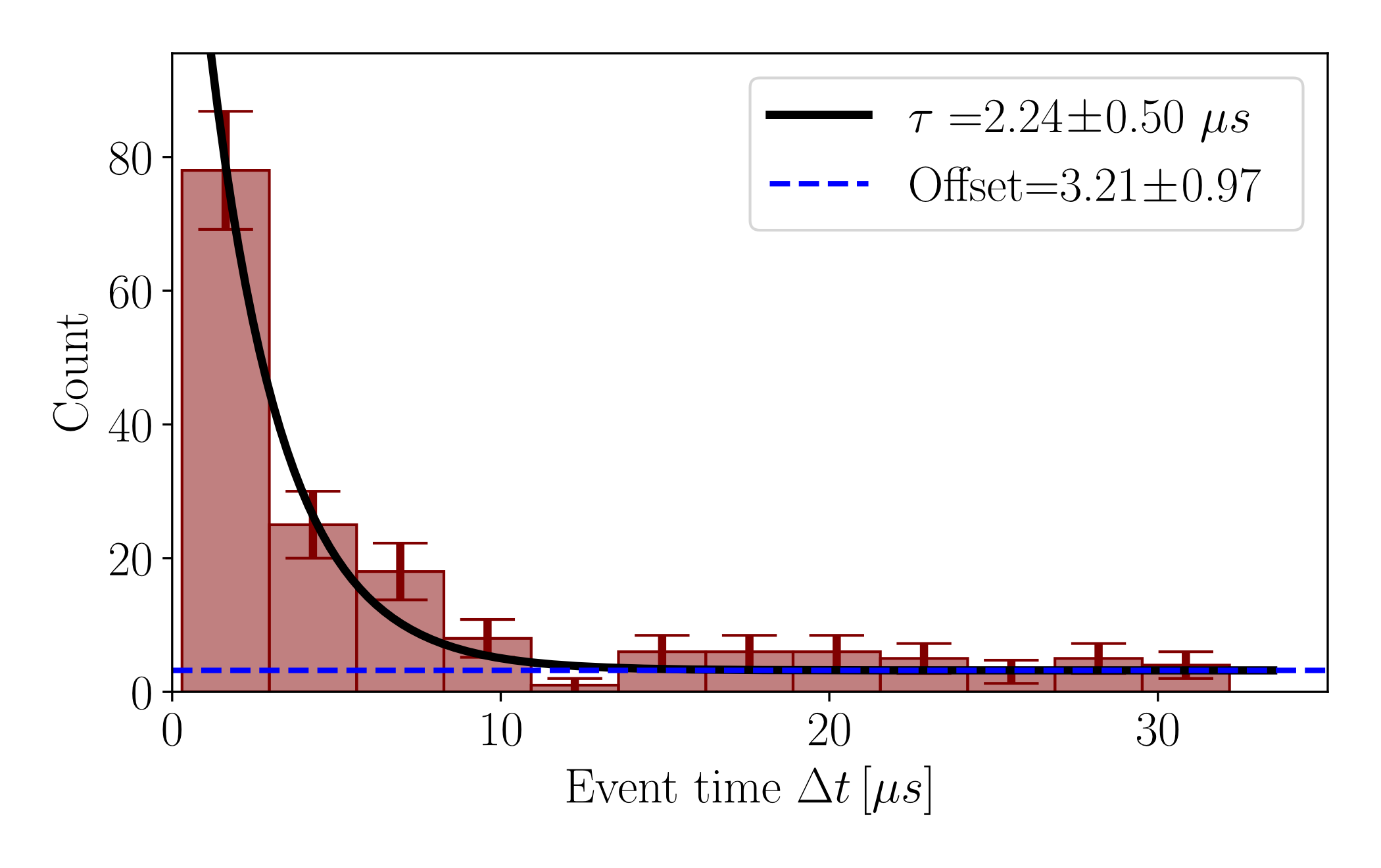}
\caption{Count of time delay from the initial trigger, $t=0$.
Error bars are obtained under a Poisson distribution.
By fitting with the exponential function $f(t)=A\exp\left(-t/\tau\right)+C$,
that is assuming signal from muon decay (black solid line) and a constant background (blue dashed line),
we obtain $\tau=2.24\pm 0.50\, \mus$ and $C=3.21\pm 0.97$.}
\label{fig:muon_decay_time_all}
\end{figure}
We assigned the time $t=0$ to each trigger and collected the subsequent signals into the histogram shown in Fig.~\ref{fig:muon_decay_time_all}.
The dataset refers to a roughly two hour period, i.e., 760 shots,
in which we measured electron beams with the average spectrum shown in Fig.~\ref{fig:electron_spectrum},
i.e, with $Q=80\,\pc$ for $E>2\,\gev$.
By fitting the histogram with the exponential function $f(t)=A\exp\left(-t/\tau\right)+C$,
we obtain the lifetime $\tau=2.24\pm 0.50\, \mus$ and a constant background $C=3.21\pm 0.97$.
The measurement of the decay time is compatible with the decay of stopped muons, i.e., $\tau_\mu=2.2\,\mus$.
Given that the total number of muon candidates recorded is $N_{tot}=165$,
and that we used 12 bins for the histogram,
the number of detected muons is $N_\mu=126\pm 12$, while the number of background events is $N_b=39 \pm 12$.
Recorded muon candidates have been verified by matching the incoming electron beam timestamp with the detection timestamp, previously synchronized.
We deem the contribution of cosmic muons negligible because
the average flux of muons from cosmic rays from all angles at sea level is about
$F\simeq 147\, \m^{-2}\s^{-1}$~\cite{particle_data_group_review_2022}, uniform in time,
i.e., the expected number of muons hitting the scintillator surface $\Sigma\simeq 0.05\,\m^2$
within a $\Delta t = 35 \,\mus$ interval after a laser shot is $\simeq 3\times10^{-4}$.
Measurements performed without operating the LPA, i.e.,
triggered by cosmic ray arrival, confirmed the expected value of flux of cosmic muons reported in~\cite{particle_data_group_review_2022}.
\begin{table}[ht]
    \centering
    \begin{tabular}{c||c|c}
        &$\tau\,\left[\mus\right]$ & $C$ \\
        \hline
        \hline
       Total & $2.24\pm 0.50$ & $3.21 \pm 0.97$ \\
        Off-axis & $2.39 \pm 0.54$ & $2.70 \pm 0.89$\\
        On-axis &$1.74\pm0.66$ & $0.58 \pm 0.32$
    \end{tabular}
    \caption{Lifetime $\tau$ and background $C$ resulting from fitting the histogram of the total, off-axis, and on-axis event, respectively.}
    \label{tab:summary_fit}
\end{table}
Using simulations, we identified the cause of the background as being generated by the photons emitted in the neutron capturing in the shielding.
After the initial burst of radiation generated by the incoming electron beam, characterized by a Bremsstrahlung spectrum very localized in time, i.e., $\Delta t \lesssim 1\,\ns$,
slower particles, such as neutrons, travel through the experimental cave and reach the concrete wall, behind the beam dump, on the $\mus$ to $\ms$ time scales. 
When neutrons are stopped in the concrete wall, they generate photons
with energies related to characteristic levels of atomic excitation~\cite{dahing_non_2014} of,
for instance, Silicon and Oxygen, the elements that account for the highest fraction of the concrete mass.
The produced $\mev$ photons hit the scintillators, scattering on its electrons and generating a signal in the PMTs.
The presence of such neutrons is confirmed by the neutron monitors inside and outside the experimental cave.
\begin{figure}[ht]
\centering
\includegraphics[width=8.6cm]{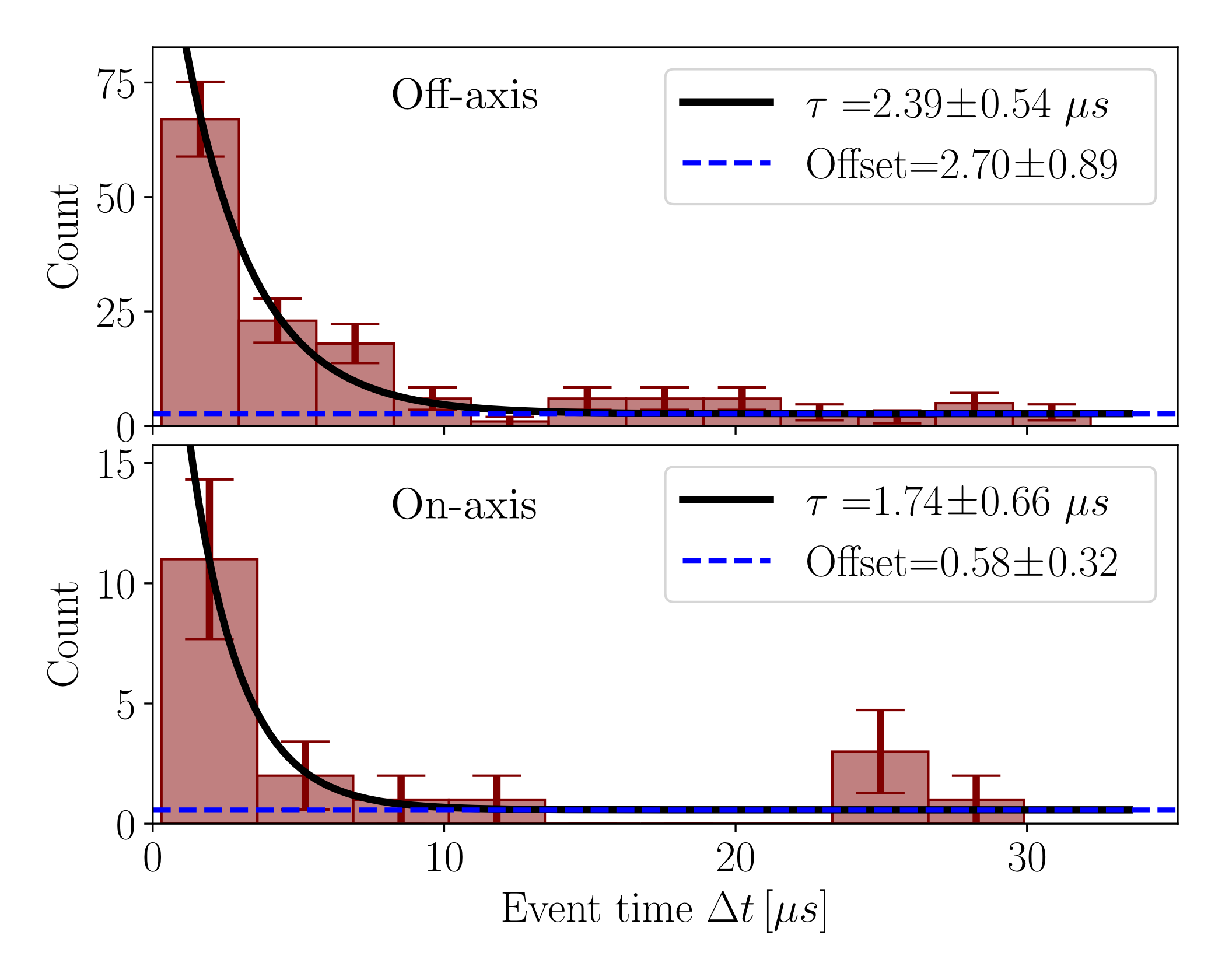}
\caption{Count of time delay from the initial trigger, $t=0$,
separated for scintillators on the beam propagation axis (bottom panel)
and off-axis (top panel).
The data included in this figure is the same as the one shown in Fig.~\ref{fig:muon_decay_time_all},
but we are presenting the muon count on and off-axis in two separate plots to highlight their difference.
Error bars are obtained under a Poisson distribution.
The exponential fits result in $\tau = 2.39 \pm 0.54 \,\mus$ and $C = 2.70 \pm 0.89$,
and $\tau=1.74\pm0.66\,\mus$ and $C=0.58 \pm 0.32$ for the off-axis and on-axis case respectively.}
\label{fig:muon_decay_time_offoncenter}
\end{figure}

In Fig.~\ref{fig:muon_decay_time_offoncenter} we separate the same muon counting dataset already presented in Fig.~\ref{fig:muon_decay_time_all} in on-axis and off-axis measurements.
These correspond to muon decay times measured with the scintillators placed on the beam axis (``on-axis'' configuration)
and $\simeq1\,\m$ to its side (``off-axis'' configuration), respectively,
both $1\,\m$ below the beam height.
The top and bottom panels show the muon counting registered from an off-axis and on-axis scintillator, respectively.
Each of the two measurements was performed in a 1-hour time interval,
and the off-axis configuration measures more muons than the on-axis one.
In particular, by repeating the analysis on the off-axis and on-axis counting histograms,
we obtain $\tau = 2.39 \pm 0.54 \,\mus$ and $C = 2.70 \pm 0.89$ and $\tau=1.74\pm0.66\,\mus$ and $C=0.58 \pm 0.32$,
respectively, from which we estimate the number of stopped muons as $N^{\text{off}}_\mu=114\pm11$ and $N^{\text{on}}_\mu=14\pm3$.
We obtain the off-axis and on-axis probabilities of a muon detection per shot per scintillator as 
$P^{\text{off}}=\left(14.4\pm1.4\right)\%$ and $P^{\text{on}}=\left(1.9\pm0.5\right)\%$ respectively,
where we used the number of shots $N^{\text{off}}_{\text{shots}}=386$ and $N^{\text{on}}_{\text{shots}}=374$.
As we will see in the next section, such a large muon count away from the beam reference axis
can be explained by taking into account the contribution of muons generated in the decay of mesons.

We have shown in this section that the analysis of the scintillator signals
revealed unambiguous detection of muons generated by the electron beams.
In the next section, we will show that computational modeling of the experiment reproduces the measurements
and identifies the source the muon count far from the beam's reference axis.

\section{Simulation results}\label{sec:simulation_results}

We performed numerical simulations of the muon production using a custom code based on the Geant4 toolkit.
Our code implements an accurate model of the beam dump and of the wall between the beam dump and the measurement room.
In addition, we included two of the high-$Z$ laser diagnostics, a wedge and power meter,
as well as the magnetic spectrometer since they are placed along the beam path, even though they do not interact with well-aligned electron beams.
The magnetic spectrometer is turned on with a field $1.08\,\text{T}$ and we verified the correct implementation by checking the bending of some reference trajectories.
We implemented a \emph{sensitive detector}, a detector in Geant4 that is a virtual screen behind the wall to record the passage of every particle.
For each particle hitting the detector, its information, e.g., phase space, detection time, and generating process, are saved.
Moreover, we implemented copies of the scintillators in order to model their response within the same simulation framework,
that is therefore fully self-consistent.
More details on the code can be found in the appendix.
\begin{figure}[ht]
\centering
\includegraphics[width=8.6cm]{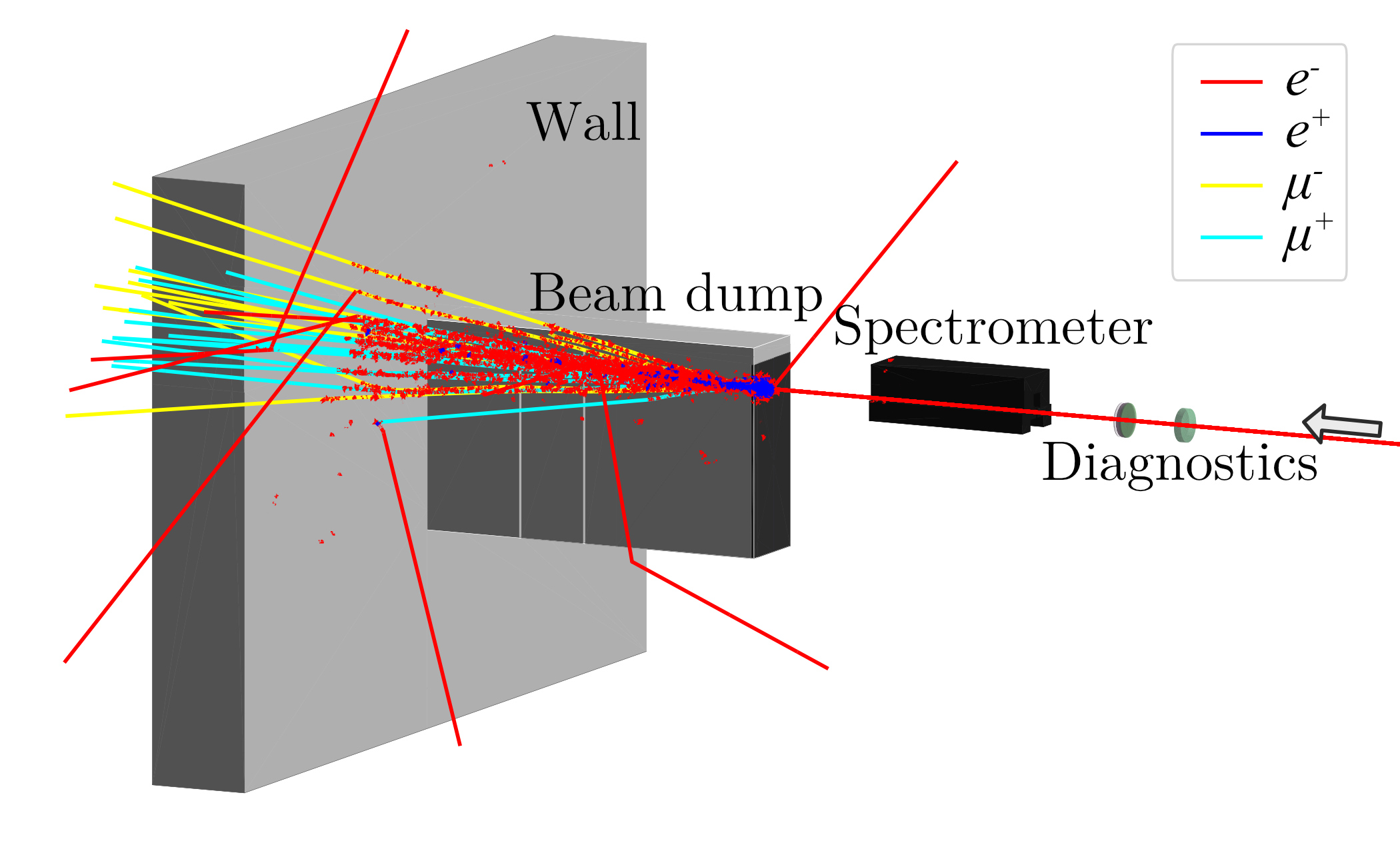}
\caption{Scheme of the simulation setup implemented in Geant4 to model the experiment.
From right to left, the arrow shows the beam direction.
The two disks are laser diagnostics with some high-$Z$ layers.
The black box is the magnetic spectrometer and, following, there is the beam dump.
On the left, the concrete wall that separates the experimental cave from the measurement room, where the scintillators are placed.
The red lines are electron trajectories, the blue ones positrons.
Positive and negative muons are respectively cyan and yellow.
We filtered neutrons and photons for the sake of visualization.}
\label{fig:BellaBeamDump_config}
\end{figure}
We initialized the electron beam according to its average spectral profile, shown in Fig.~\ref{fig:electron_spectrum},
limited to energies $E>2\,\gev$,
i.e., with a total charge $Q=80\,\pc$,
and with a \emph{rms} divergence $0.1\,\mrad$.
The \emph{rms} divergence was obtained by Particle-in-Cell (PIC) simulations and further confirmed in the measurements, which showed a divergence in the range $0.1-0.5\,\mrad$.
We point out that the exact value does not affect the overall description of the process,
as long as it remains small, i.e., $<1\,\mrad$,
since emission angle of secondary particles is dominated by the electron scattering, with a typical angle $\geq 100 \,\mrad$.
Fig.~\ref{fig:BellaBeamDump_config} shows the simulation setup,
including, from right to left, following the beam path, the two laser diagnostics,
the magnetic spectrometer, the electron beam dump, and the large concrete wall behind the electron beam dump.
The width and height of the wall coincide with the spatial extent of the Geant4 geometry,
therefore it is not possible for a particle to travel around the wall and reach the detectors, not shown in the figure, behind the wall itself.
In Fig.~\ref{fig:BellaBeamDump_config}, red trajectories are the electrons,
including the primary ones coming from the direction of the arrow, while blue trajectories represent the positrons.
The cyan and yellow trajectories are, respectively, $\mu^+$ and $\mu^-$, which are created
in the early lead region of the dump and propagate through it, reaching the measurement room.
For the sake of visualization, we removed photon and neutron trajectories from the picture.
As explained in Sec.~\ref{sec:theory}, we expect the majority of high energy photons to be produced in the beam direction
via Bremsstrahlung.
Neutrons are produced in the high-$Z$ layers of the beam dump and emitted isotropically in space.
Electrons with energies ranging from $\kev$ to $\mev$ are extracted via ionization by the muon passage and, in the majority of cases, quickly reabsorbed.
Electrons extracted close to the back of the wall can have enough energy to escape and propagate into the measurement room.
Such electrons are represented in Fig.~\ref{fig:BellaBeamDump_config} by the red trajectories that generate from the muons ones.
\begin{figure}[ht]
\centering
\includegraphics[width=8.6cm]{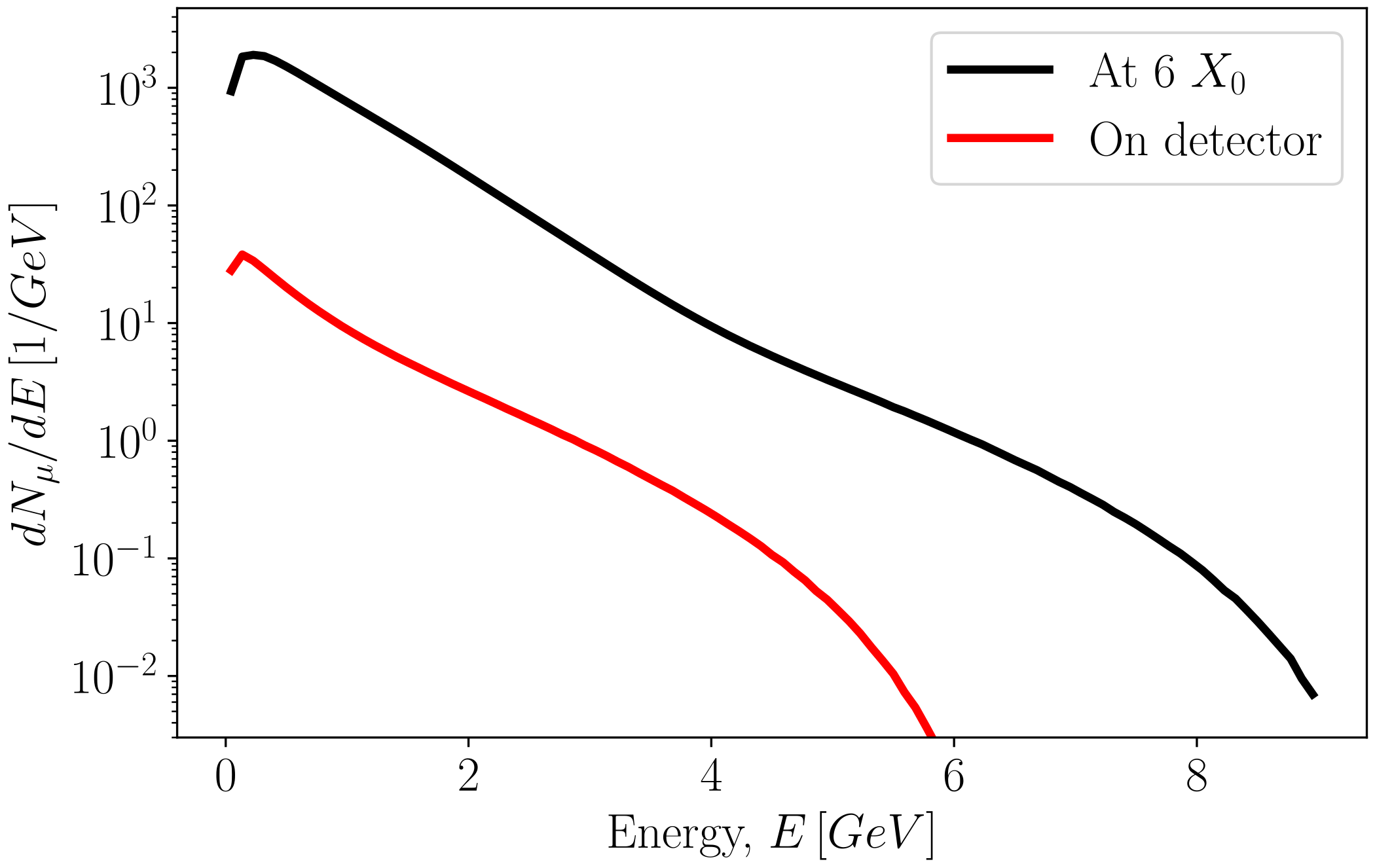}
\caption{Muon energy spectra obtained via Geant4 simulations of a perfectly aligned electron beam
with energy shown in Fig.~\ref{fig:electron_spectrum} and $80\,\pc$ charge.
The black line is the energy spectrum sampled 6 radiation lengths into the dump.
The red line is the spectrum on the virtual screen behind the external wall.
The difference between the spectra is due to the ionization losses,
which roughly account for $E_L \lesssim 4 \,\gev$.
}
\label{fig:energy_losses}
\end{figure}

We show in Fig.~\ref{fig:energy_losses} energy spectra of muons, resulting from Geant4 simulations, sampled at $6\,X_0$ inside the electron beam dump,
i.e., $\simeq 5\,\cm$ inside the lead block (black line)
and on the detector in the measurement room (red line).
Muons predominantly lose energy via ionization when propagating through dense materials.
Traversing the electron beam dump and the wall results in $E_L\lesssim 4\,\gev$ muon energy loss before the detection point.
This can be verified by integrating the equation of motion of a muon
moving in a straight trajectory through the shielding configuration,
using the stopping power tabulated in Ref.~\cite{groom_muon_2001}.
In a broad energy range $1\,\gev\leq E\leq10\,\gev$ the stopping power is substantially energy-independent,
thus the calculation applies to all the generated muons.
\begin{figure}[ht]
\centering
\includegraphics[width=8.6cm]{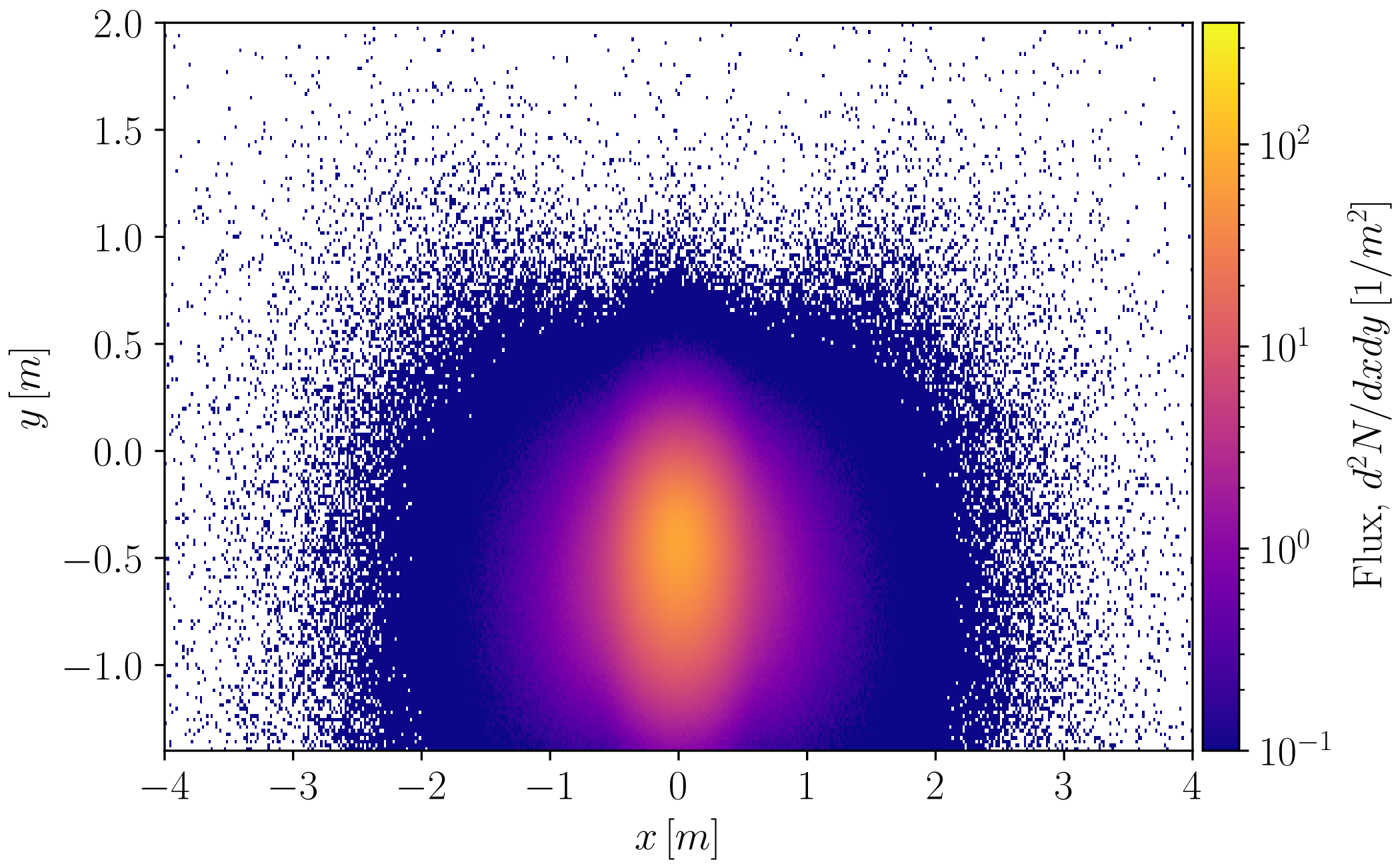}
\caption{Muon flux behind the concrete wall computed using Geant4 for a perfectly aligned beam.
The calculation assumes an initial $Q=80\,\pc$ electron beam perfectly aligned with the ideal beam's reference axis,
i.e., it does not interact with metal parts in the laser diagnostics and the magnetic spectrometer chassis.
The muon flux at $x=0,\,y=-0.45\,\m$ is $F=73.5\pm9.1\,\m^{-2}\text{shot}^{-1}$.
An inspection of the muons shows that all of them are generated via Bethe-Heitler.
The detection plane is $\simeq 4 \,\m$ downstream from the muon source.}
\label{fig:muon_flux_pair_center}
\end{figure}

Figure~\ref{fig:muon_flux_pair_center} shows the expected flux of muons obtained from an
$80\,\pc$ electron beam with the initial energy spectrum shown in Fig.~\ref{fig:electron_spectrum}.
The beam center before the spectrometer is in $x=0,\,y=0$, and the axis $y=-1.4\,\m$ corresponds to the floor.
Due to the bending of the incoming electron beam in the magnetic field of the spectrometer,
the average vertical position of the muons on the detector is $y=-0.45\,\m$.
The muon flux at $x=0,\,y=-0.45\,\m$ is $F=73.5\pm9.1\,\m^{-2}\text{shot}^{-1}$.
Simulations show that the number of muons with an angle from the beam axis $\theta \leq \theta_0= 100\,\mrad$
impinging on a surface $\Sigma=2\times 2\,\m^2$ placed in front of the source
is $N_{\mu}^{LPA}\gtrsim 20 \,\text{shot}^{-1}$.
These muons are, for instance, the ones of interest for muography applications along the earth's surface plane.
In comparison, based on the approximated description of the muon flux from cosmic rays,
we can estimate the arrival rate of such muons on the same surface and within the same acceptance
as $N_{\mu}^{CR} \simeq 147\sin^3(\theta_0)\Sigma\simeq 0.5\, s^{-1}$.
Assuming LPAs generating $\gtrsim 100\,\pc$ beams and operating at $1\,\hz$, accessible to most currently-available high-power laser systems,
laser-plasma generated electron beams provide more than
40 times the muons than cosmic rays along the plane of the earth's surface.
Development of high-repetition-rate laser drivers can boost this factor by orders of magnitude.
The plot shows a negligible flux of muons outside
a cone of narrow aperture $\sigma_{\theta}\simeq 100\,\mrad$,
which agrees with the expectation of muons generated via Bethe-Heitler.
\begin{figure}[ht]
\centering
\includegraphics[width=8.6cm]{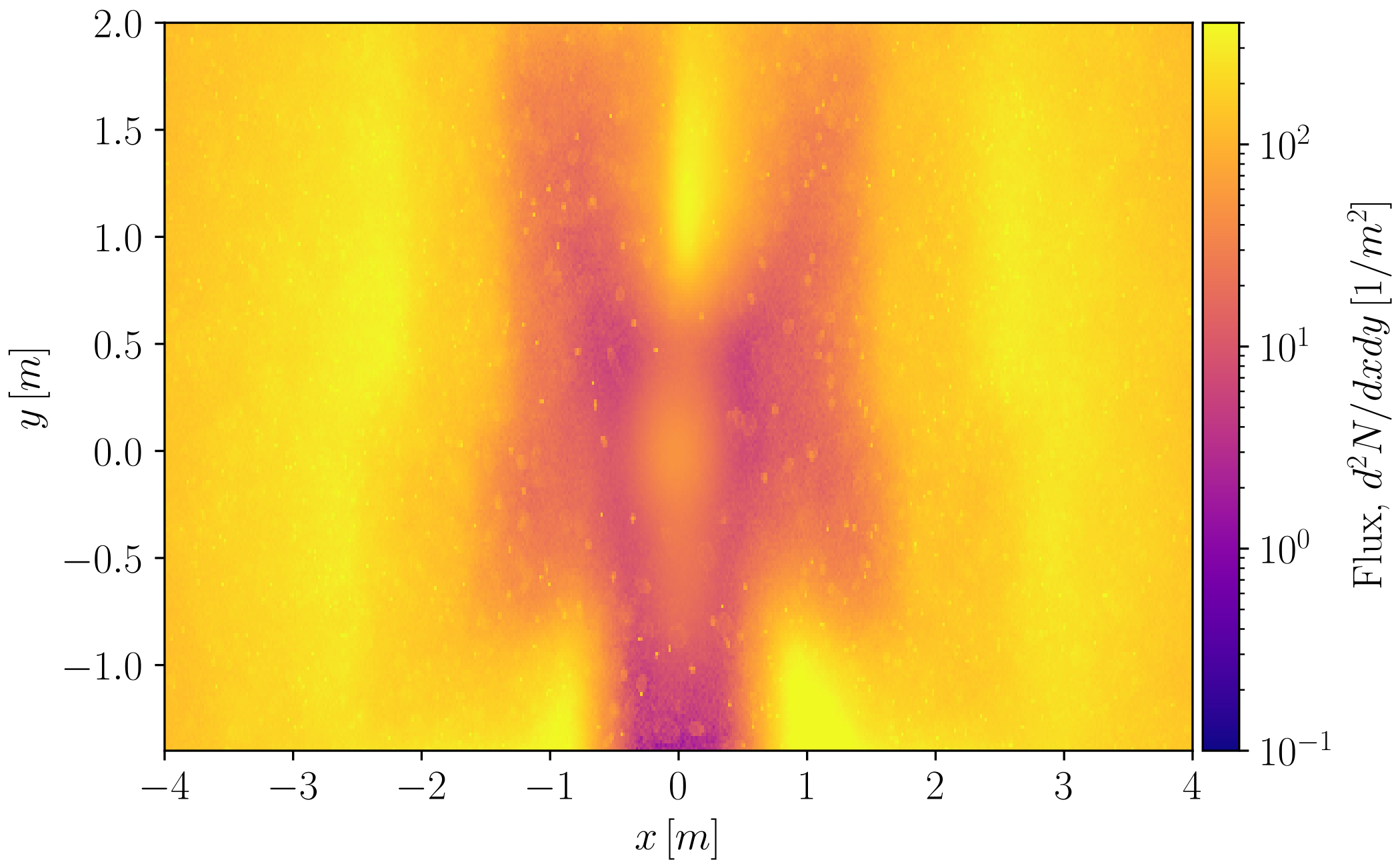}
\caption{Muon flux behind the concrete wall computed using Geant4 for a misaligned electron beam.
The calculation assumes an initial $Q=80\,\pc$ electron beam, with an initial misalignment $1.5\,\mrad$ from the beam's reference propagation axis.
The shift is sufficient for the electron beam to interact with the metal in laser diagnostics and the magnetic spectrometer chassis.
The central region, close to the beam's reference axis,
is mainly composed of muons generated via Bethe-Heitler.
The flux at $x=0$, $y=-0.45\,\m$ is $F=24.9\pm9.1\,\m^{-2}\text{shot}^{-1}$ and its characteristic vertically oblong shape around this point resembles the well-aligned beam case in Fig~\ref{fig:muon_flux_pair_center}.
Muons generated in meson decay contribute to a quasi-isotropic flux, visible on the detector far from the beam's reference axis.
The detection plane is $\simeq 4 \,\m$ downstream from the muon source.}
\label{fig:muon_flux_misaligned}
\end{figure}
In order to explain the detection of muons away from the beam's reference axis,
we need to take into account the pointing fluctuations of the electron beam,
characterized during the experiment with an \emph{rms} value $\sigma_{\theta}\simeq 3\,\mrad$.
Despite this value is negligible compared to the aperture of the photon and muon emission cone,
it is enough for the electron beam to interact with
the laser diagnostics placed before the magnetic spectrometer, which have an angular acceptance of $1.2\,\mrad$,
and with the magnetic spectrometer itself.
Since all these elements are made of high-$Z$ materials, i.e., aluminum, fused silica, and steel,
a beam passing through them produces muons via Bethe-Heitler process and mesons from photoproduction,
as illustrated in Sec.~\ref{sec:theory}.
Mesons do not encounter the same amount of material as when they are produced deep inside the electron beam dump and
therefore they are not reabsorbed, propagating quasi-isotropically in the experimental cave.
Once these mesons decay, the low energy muons that are produced do not need to traverse the electron beam dump
and have enough energy to propagate through the concrete wall and be detected by the scintillators.
This is demonstrated in Fig.~\ref{fig:muon_flux_misaligned}, where we show the simulated flux of muons per beam,
assuming an initial beam deviation of $1.5\,\mrad$ from the reference axis.
The central shade in the picture resembles the shape of the magnetic spectrometer and the beam dump,
which demonstrates that some mesons are produced and decay into muons before the spectrometer,
resulting in a radiography of the two elements onto the virtual detector.
Inside the shaded area, around the point $x=0,\,y=-0.45\,\m$, we can see the collimated contribution to the flux of the muons generated via Bethe-Heitler, similar to the well-aligned case shown in Fig.~\ref{fig:muon_flux_pair_center}.
The flux at $x=0$, $y=-0.45\,\m$ is $F=24.9\pm9.1\,\m^{-2}\text{shot}^{-1}$.
We notice an elongated region of high muon flux on the $x=0$ axis, for $y\gtrsim1\,\m$.
Simulations identify these muons as being generated via Bethe-Heitler.
The interaction of the electron beam with the high-Z elements before the magnetic spectrometer produces positive particles,
such as positrons and antimuons, which are bent upward by the magnetic field.
Due to the geometry of our shielding, as shown in Fig.~\ref{fig:BellaBeamDump_config}, these positive particles only traverse a small fraction of the beam dump, which results in a higher muon flux.
Outside of the shaded area, muons from meson decay contribute to a uniform, quasi-isotropic flux.
The lack of filtering power provided by the electron beam dump explains the higher background noise recorded in off-axis measurements.
Photons generated in the interaction and in the neutron capturing inside the shielding,
can traverse the wall away from the beam dump,
scattering on the electrons of the scintillators and triggering signal.
\begin{figure}[ht]
\centering
\includegraphics[width=8.6cm]{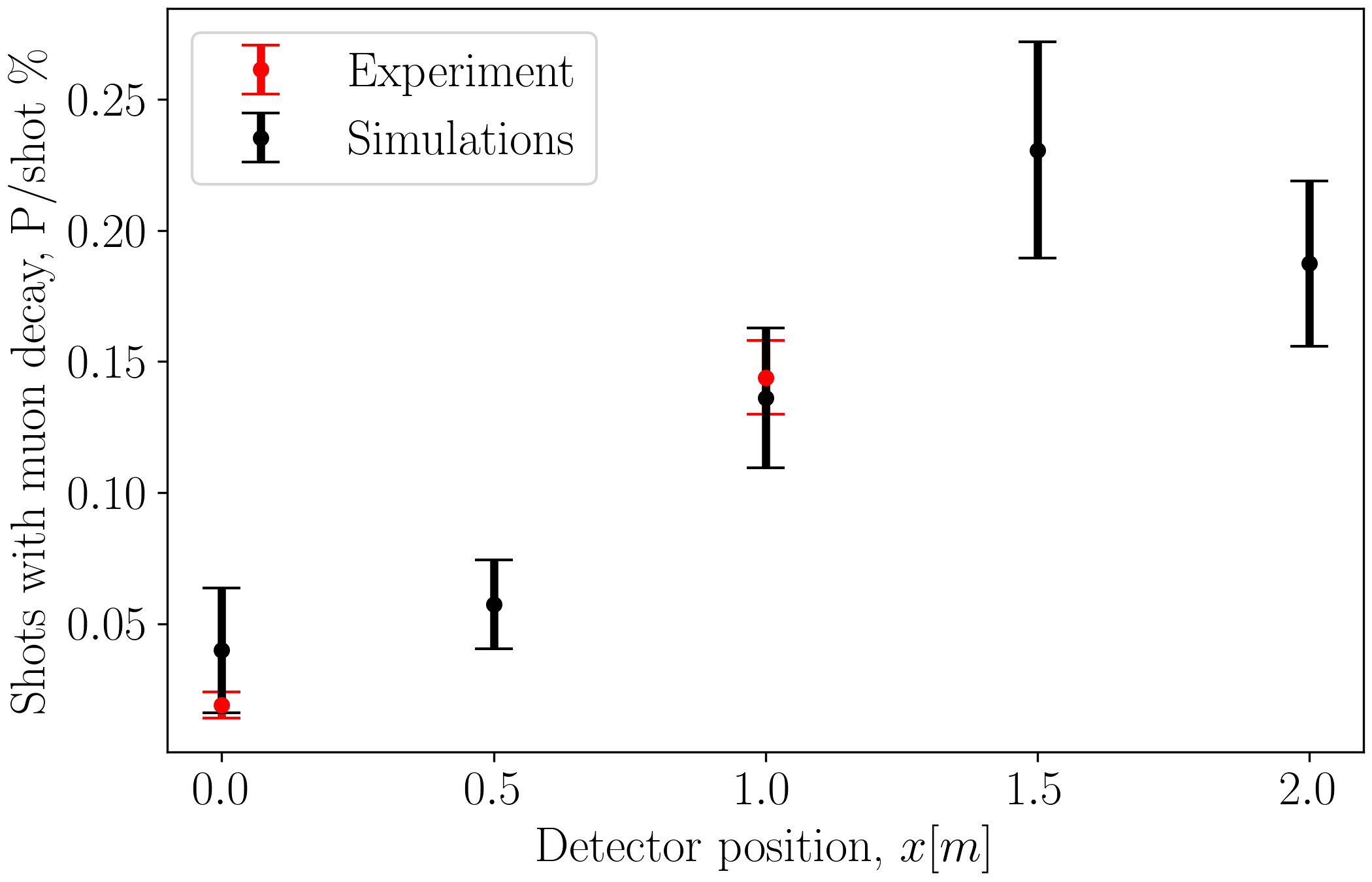}
\caption{Simulated percentage of LPA shots with a positive muon count in a scintillator
in function of the horizontal position of the scintillator.
Results are obtained using a model of the scintillator implemented in Geant4.
The probabilities at $x=0$ and $x=1\,\m$ agree well with the values measured in the experiment.
Scintillators are placed $1\,\m$ below the beam height.
Simulation results are obtained by (weighted) averaging of two simulation scenarios: the case of well-aligned electron beams (with $<$1.2 mrad pointing deviation from ideal reference axis, observed during 30\% of the LPA shots) and misaligned electron beams at $>$1.2 mrad pointing (70\% of the LPA shots).}
\label{fig:count_probability_sim}
\end{figure}

The scintillator models implemented in the code allow us to directly count the muons that stop and decay into them,
reproducing the signal we obtained in the experiment.
In order to approximate the effect of the pointing fluctuations of the electron beams,
we consider the results obtained by simulating a linear combination of well-aligned and misaligned beams.
Given that the angular fluctuations were characterized during the experiment
as being gaussian with $\sigma_\theta\simeq 3\,\mrad$,
the number of beams contained within $-1.2\,\mrad \leq \theta \leq 1.2\,\mrad$,
that is the acceptance of the optical elements before the dump, determining a well-aligned beam, is $\sim 30\%$.
We can see in Fig.~\ref{fig:count_probability_sim} the simulated probability of counting a muon per LPA beam in detectors
placed on the $y=-1\,\m$ axis, matching their placement in the experiment, as a function of the detector horizontal position, for a combination of $30\%$ aligned beams and $70\%$ misaligned beams.

The results from the simulations agree quite well with the experiment.
We computed a probability of a muon stopping in a scintillator $P^{\text{off}}=(13.6\pm2.7)\%$ for a scintillator placed $1\,\m$ to the side of the deflected reference axis, and $P^{\text{on}}=(4.0\pm2.4)\%$ for a scintillator placed in the beam's reference axis.
Recalling the results shown in Sec.~\ref{sec:experimental_results}, the measured values for those probabilities are
$P^{\text{off}}=\left(14.4\pm1.4\right)\%$ and $P^{\text{on}}=\left(1.9\pm0.5\right)\%$ respectively.

Scintillator-based detectors cannot measure the muon flux directly.
The majority of muons crosses them within a short temporal window $\Delta t \lesssim 1\,\ns$
following the LPA beams, together with secondary particles.
This prevents a direct muon counting because the PMT is saturated and it is not possible to distinguish between each different charged particles triggering a signal.
However, the agreement between simulations and measurements on the number of stopped muons both on- and off-axis, i.e., the low energy fraction of the incoming ones,
provides insight into the produced flux, which simulations predict being on the order of $F\gtrsim 10\, \m^{-2}\text{shot}^{-1}$ for the electron beams produced in our experimental run.
We have shown that the flux is the result of two spatially separated contributions:
a central region, around the beam axis, of directional, high-energy muons generated via pair production
and an off-axis region mainly composed by lower energy non-directional muons generated via meson decay.
\begin{figure}[ht]
\centering
\includegraphics[width=8.6cm]{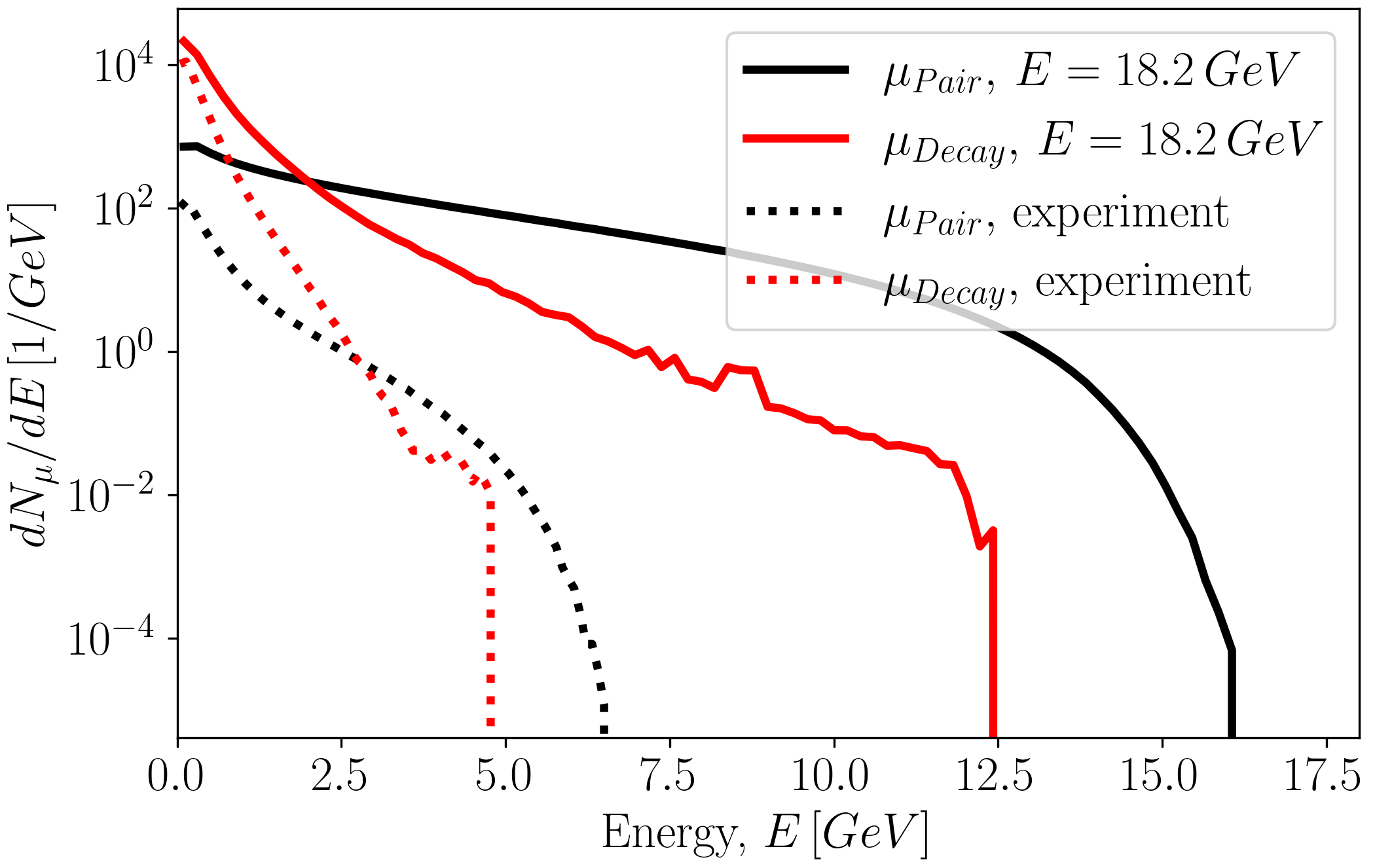}
\caption{Spectra of muons behind the wall, obtained using Geant4, integrated over the whole surface
and differentiated by the muon generating process.
The black lines show the spectrum of pair-produced muons, the red lines show muons generated via meson decay.
The solid lines refer to a monochromatic, $18.2\,\gev$, $26\,\pc$ electron beam obtained in simulations of staging LPAs
using two $19\,\joule$ laser pulses, with BELLA laser parameters.
The dotted lines refer to the spectra obtained using the experimental electron beam with spectrum as in Fig.~\ref{fig:electron_spectrum}.
Increasing the electron beam energy improves the source quality by producing a higher yield of directional and high energy muons.
}
\label{fig:spectra_pair_decay}
\end{figure}

The present work shows that LPAs provide unique advantages
as a source of directional muon beams.
Their ultra-high accelerating gradients produce electron beams
within very compact, i.e., 10s of cm scales, acceleration lengths.
These are orders of magnitude smaller compared to the storage rings used to accelerate protons that produce muons via the pion decay channels.
Such electron beams possess
an outstanding capability of producing directional muon beams.
The performance of the LPA source can be substantially improved by increasing the initial electron energy
due to a conjunction of multiple effects.
The properties of the Bremsstrahlung cascade are such that the yield of particles from pair production for an optimized target
scales roughly linearly with the initial beam energy~\cite{chao_handbook_2013},
although the scaling could be either sub- or super-linear depending on the target geometry
and the exact energy range~\cite{calvin_laser-driven_2023}.
Moreover, higher initial energies result in a more collimated muon beam, i.e., higher flux,
and a higher maximum muon energy.
We compared simulations of the muons produced by the electron beam obtained in our experiment
to the ones produced by a higher energy beam achievable, for instance,
by staging two LPA sections.
We used the PIC code INF\&RNO~\cite{benedetti_efficient_2010, benedetti_accurate_2017} to obtain the final electron beam distribution.
As an example, we considered a realistic setup based on the current $40\,\joule$ BELLA facility where each stage is powered by a $19\,\joule$ laser beam.
PIC simulations produced a $18.2\,\gev$ beam with $26\,\pc$ of charge and small energy spread, which we used as input in our Geant4 code to understand how the muon production scales at higher energy.
Fig.~\ref{fig:spectra_pair_decay} shows a comparison between the simulated muon spectra, integrated over the entire wall surface,
resulting from the interaction with the shielding configuration of the average $80\,\pc$
beam obtained in our experiment, labeled ``experiment'', and of the $26\,\pc$ monochromatic beam with $E=18.2\,\gev$
produced in simulations of staging acceleration,
respectively.
The plot demonstrates that by increasing the source energy, the overall quality of the muon source improves.
Muons produced via pair production extend to higher energies and,
even accounting for the energy losses induced by the shielding,
they possess the strong penetrating power required for imaging applications.

These results motivate us to design future experiments and applications based on $E>10\,\gev$
electron beams.
While for the present work we relied on the electron-to-muon conversion offered by the shielding,
it is possible to design optimized converter targets, as shown in Fig.~\ref{fig:material_scan},
where the muon flux can be maximized and the energy losses minimized.
The optimal energy of the electron beam is determined by the imaging sample considered,
due to the aforementioned effect of ionization losses.
We note that, in the configuration used here, scintillators are not suitable for imaging applications.
Such applications rely on tracking individual muon trajectories before and after the sample and typically demand muons with energies
anywhere from $\sim 100\,\mev$ to beyond $1\,\tev$, the exact requirement being set by the desired penetration depth.
Detectors used for tracking should therefore be capable of registering every muon that arrives, irrespective of its energy.
In our experiment, however, we characterized the muon source by counting delayed signals in the scintillator panels, a method that selects only a small fraction of low-energy muons.
Nonetheless, the results suggest that LPAs could provide a practical muon source for imaging applications requiring penetration depths achievable with muons below a few tens of $\gev$.

With the Geant4-based simulation tool that we developed, we can investigate the properties of the resulting beam and design an appropriate detection scheme to pick up signals from the high-energy muons before and after the imaging sample.
It is also possible to equip our code with imaging reconstruction algorithms that enable start-to-end simulations, i.e., from the electron beam to the final imaging result,
which would let us tailor the design of a converter target and the required shielding.

\section{Conclusions}\label{sec:conclusions}

In this paper, we presented the implementation of a high-energy and directional muon source
based on laser-plasma accelerated electron beams obtained using the BELLA PW laser.
The electron beams were generated via ionization injection of an unlocalized dopant
and accelerated using a $21\,\joule$ laser pulse,
which yielded beams with broad energy spectra reaching up to $8\,\gev$,
low divergence, and pointing fluctuations of $3\,\mrad$~\cite{picksley_matched_2024}.
When such beams impinged on the electron beam dump or grazed other high-$Z$ elements along their path,
muons were generated via pair production and meson decay.
We identified $N=126\pm12$ muon detections during two hours operation,
sampled from the low energy part of the generated beam that stopped into the scintillators,
and we characterized the resulting muon flux via simulations and experiments,
noting that the two production mechanisms could be spatially distinguished.
Muons from pair production are characterized by a high-energy,
up to the energy of the initial electrons, a low-divergence, and a strong directionality along the incoming beam path.
These characteristics make them very suitable for imaging applications as the illumination of a target can be optimized
to maximize their flux and their tracking in and out from the target itself.
The energy and number of pair produced muons scales with the initial electron beam energy and charge
and can be tuned to fit the application requirements.

Muons from decay of mesons make up the majority of muons near the converter.
They are characterized by a quasi-isotropic angular distribution and by a sharp cutoff at high energies,
therefore their contribution to the muon flux in the forward direction is very weak.
It is not desirable to rely on them in applications that require high fluxes at a distance and strong penetrating power.
In our experiment and simulations we showed that muons generated via meson decay
can be detected away from the beam path, where they are not filtered by the electron beam dump.

LPA-accelerated electron beams are generated over compact acceleration lengths and possess outstanding properties to generate
muon yields orders of magnitude higher than what is naturally available with cosmic rays.
In particular, while the verage number of cosmic muons through a meter-scale imaging setup parallel to the earth's surface plane
with an aperture $\theta=100\,\mrad$ is $N_{\mu}^{CR}\leq 1\,\s^{-1}$, 
LPA beams with energies and charges in the range, respectively, $E \sim \left(1-10\right)\,\gev$ and $Q\sim \left(10-500\right)\pc$
produce $N_{\mu}^{LPA}\gtrsim 10^2\,\text{shot}^{-1}$ through the same imaging setup.
Near future LPA and laser technology developments, e.g., staged acceleration and pulse delivery at the kHz repetition rate,
can boost this property by orders of magnitude.
Using the same simulation tools we used to understand the experimental results,
we can design future muography application based on high-energy electron beams and
detectors suitable for muon scattering image reconstruction.

This experiment proves the feasibility of a muon source based on LPA-generated electron beams.
It is an important result to establish the readiness of the LPA technology for applications
and to motivate the development of high-repetition-rate and transportable laser systems,
which could push applications like muography to unprecedented capabilities,
offering muon yields orders of magnitude higher than cosmic rays.

\begin{acknowledgments}
The authors would like to thank Adam Davis, Brian Dorney, Francesca Kukral, Hans Malik, and Scott Wolin
of Northrop Grumman Corporation, Ryan Heller of the Lawrence Berkeley National Laboratory for the insightful discussions, and Zachary Eisentraut, Mark Kirkpatrick, Federico Mazzini, Nathan Ybarrolaza, Derrick McGrew, Paul Centeno, Teo Maldonado Mancuso, Art Magana, Joe Riley, Mackinley Kath, and Chetanya Jain of the Lawrence Berkeley National Laboratory for technical support on experiments. 
D.T. is grateful to Sarah Schr\"oder and Stepan Bulanov for the enlightening conversations.
This work was supported by the Defense Advanced Research Projects Agency (DARPA),
the Director, Office of Science, Office of High Energy Physics,
of the U.S. Department of Energy under Contract No. DE-AC02-05CH11231,
and used the computational facilities at the National Energy Research Scientific Computing Center (NERSC).
\end{acknowledgments}

\appendix*

\section{Modeling the experiments with a Geant4-based code}

We performed simulations of the experimental campaign with a custom application developed using Geant4.
The code defines the geometry of the experimental chamber retaining shapes and materials of the included elements.
The implementation includes two laser diagnostics, a wedge and a power meter, the magnetic spectrometer,
the beam dump with all its internal layers and the wall.
The electron beam can be initialized freely using macros,
and its energy spectrum, initial position and direction and its divergence can be chosen arbitrarily.
The code uses the predefined \emph{QGSP\_BERT\_EMZ} physics package.
It includes a high precision model for neutron propagation, that we used to understand background noise,
and for the electromagnetic interaction.
In order for the code to account for the muon pair production, 
the process has to be manually activated via a macro command,
because it is not included in the pre-defined physics list by default.
The code uses biasing to improve the statistical significance of the simulations.
In particular, we rely on the splitting of muon pairs and pion decays to increase the number of produced muons.
During our time using and benchmarking the code we updated several Geant4 versions,
but the results presented in this paper are obtained with Geant4 \texttt{v11.2}.
While the version \texttt{v11.0} returns consistent values,
we noticed during our investigation that version \texttt{v11.1} does not yield the same 
results for muon production. The issue was addressed in subsequent releases 
(see Geant4 \texttt{v11.2} Release Notes).

The code is equipped with several diagnostics,
including sensitive detectors in the dump that track the development of the Bremsstrahlung cascade,
a sensitive detector behind the wall that captures all the particles leaving the chamber
and a model of the scintillators.
We save the phase space and some relevant information about particles that cross them
that generate most of the plots shown in the paper.
The scintillators are constructed using the same material and size as the real detectors
and they are programmed to return particles that stop within their boundaries,
from which we can extract the number of muons detected.

\bibliography{BellaMu}
\end{document}